\documentclass[runningheads]{llncs}

\usepackage[T1]{fontenc}
\usepackage{graphicx}
\usepackage{amsmath}
\usepackage{amssymb}
\usepackage{booktabs}
\usepackage{siunitx}
\usepackage{subcaption}

\newcommand{\fdg}{\ensuremath{\mathrm{[^{18}F]FDG}}}

\DeclareMathOperator*{\argmin}{argmin}

\DeclareSIUnit{\minutes}{minutes}
\DeclareSIUnit{\weeks}{weeks}
\DeclareSIUnit{\hours}{hours}

\begin{document}

\title{Physics-Informed Deep Learning for Improved Input Function Estimation in
Motion-Blurred Dynamic [$^{18}$F]FDG PET Images\thanks{Preprint: accepted to PRIME @
MICCAI 2025. This is the submitted (pre-review) version.}}
\titlerunning{Physics-Informed Deep Learning Based Input Function Prediction}

\author{Christian~Salomonsen\inst{1}\orcidID{0009-0007-4958-4544} \and
Kristoffer~K.~Wickstrøm\inst{1}\orcidID{0000-0003-1395-7154} \and
Samuel~Kuttner\inst{1,2}\orcidID{0000-0001-7747-9003} \and
Elisabeth~Wetzer\inst{1}\orcidID{0000-0002-0544-8272}}

\authorrunning{C. Salomonsen \emph{et al.}}

\institute{Department of Physics and Technology, UiT The Arctic University of Norway, Tromsø, Norway \and PET Imaging center, University Hospital of North Norway, Tromsø, Norway}

\maketitle

\begin{abstract} 

Kinetic modeling enables \textit{in vivo} quantification of tracer uptake and
glucose metabolism in [$^{18}$F]Fluorodeoxyglucose (\fdg{}) dynamic positron
emission tomography (dPET) imaging of mice. However, kinetic modeling requires
the accurate determination of the arterial input function (AIF) during imaging,
which is time-consuming and invasive. Recent studies have shown the efficacy of
using deep learning to directly predict the input function, surpassing
established methods such as the image-derived input function (IDIF). In this
work, we trained a physics-informed deep learning-based input function
prediction model (PIDLIF) to estimate the AIF directly from the PET images,
incorporating a kinetic modeling loss during training. The proposed method uses
a two-tissue compartment model over two regions, the myocardium and brain of the
mice, and is trained on a dataset of 70 \fdg{} dPET images of mice accompanied
by the measured AIF during imaging. The proposed method had comparable
performance to the network without a physics-informed loss, and when sudden
movement causing blurring in the images was simulated, the PIDLIF model
maintained high performance in severe cases of image degradation. The proposed
physics-informed method exhibits an improved robustness that is promoted by
physically constraining the problem, enforcing consistency for
out-of-distribution samples. In conclusion, the PIDLIF model offers insight into
the effects of leveraging physiological distribution mechanics in mice to guide
a deep learning-based AIF prediction network in images with severe degradation
as a result of blurring due to movement during imaging.

\end{abstract}

\section{Introduction}

\fdg{} Positron emission tomography (PET) is a common
medical imaging modality that produces spatial representations of a
radiopharmaceutical being metabolized by tissue and cells with very high
sensitivity. PET images therefore provide insight to what tissue and cells are
active, and provide an empirical method for detecting malignant tissue such as
cancerous tumours---represented by bright spots in the PET image---or monitoring
cancer therapy. These \textit{static} PET images are traditionally constructed
from averaging the gamma ray count over time, when the radiotracer uptake have
stabilized long after radiotracer injection. 

\begin{figure}[t]
    \centering
    \begin{subfigure}[t]{0.45\columnwidth}
        \centering
        \includegraphics[width=0.6\linewidth]{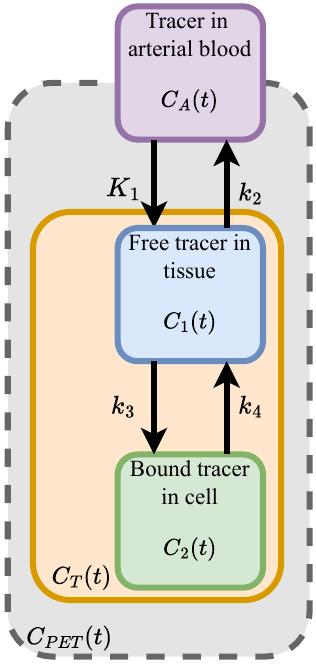}
        \caption{Reversible two-tissue compartment model.}
        \label{fig:2TC}
    \end{subfigure}
    \begin{subfigure}[t]{0.45\columnwidth}
        \includegraphics[width=\linewidth]{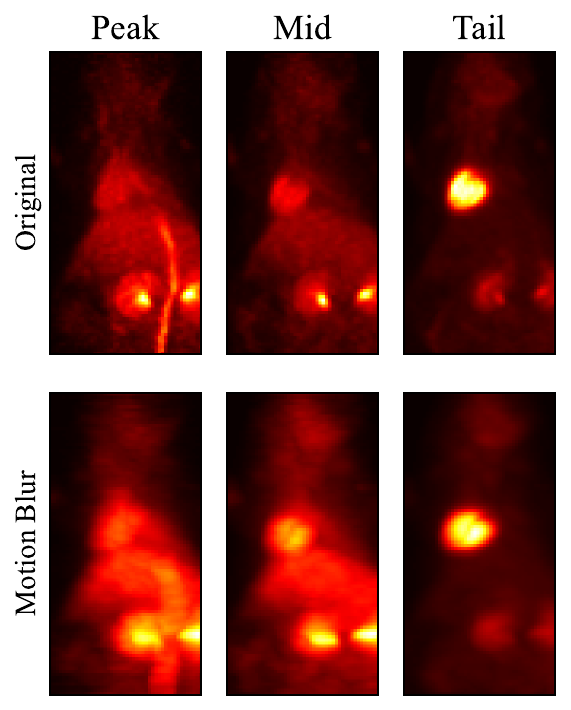}
        \caption{Simulated motion.}
        \label{fig:motionblur-example}
    \end{subfigure}
    \caption{Compartment model (a) for kinetic modeling with \fdg{} PET with
    rate constants describing rate of transfer between compartments. For this
    system, the metabolic rate of glucose is proportional to the net influx
    rate, \(K_i \equiv K_1 \cdot k_3 / (k_2 + k_3)\), where rate constants
    $K_1,\ k_2$, and $k_3$ are derived from the compartment model. Example of
    artificial motion blur (b) split over tracer uptake phases. A motion blur
    kernel of size \numproduct{15x15x15} are convolved with the image to
    simulate movement in a random direction during imaging. The columns
    correspond with the first 25 (peak), intermediate 9 (mid), and last 8 (tail)
    time-frames, respectively. The images are averaged over time within each
    respective segment, resulting in crisper appearence than the actual input
    data.}
    \label{fig:intro}
\end{figure}

While static PET is useful for its applications, the nature of the modality
limits its usability to other downstream tasks, such as in developing new drugs,
tracers, diagnostic procedures and disease
therapies~\cite{hicks2006pet,yao2012small,cunha2014preclinical}. The alternative
to static PET is dynamic PET (dPET), where radiation detections from the full
imaging duration are averaged into smaller temporal bins, resulting in a spatial
time-series representation of the entire tracer uptake process. Thus, providing
\textit{in vivo} visualizations of the tracer uptake captured in time-activity
curves (TACs) that represent the temporal variability of tracer binding to
proteins, being in a free state in plasma, or being metabolized. Further
analysis is then done using tracer kinetic modeling, which is a tool that
allows for estimating the uptake rates between plasma, tissue, and
cells~\cite{Gunn2001b}. Kinetic modeling typically builds on an assumed
mathematical model for the particular organ or tissue, usually in the form of
compartment modeling, and uses the captured dPET spatial time-series and the
plasma blood concentration \cite{khalil2017basicPET,Alf2013e}, also known as the
arterial input function (AIF). While dPET imaging is practically straight
forward, acquiring an accurate representation of the AIF is not trivial;
arterial cannulation during image acquisition is considered the gold-standard,
but requires invasive surgical cannulation, which is time-consuming, complex,
and also restricted in the amount of blood available to not alter the physiology
of the mice~\cite{Laforest2005}. In these studies, arterial blood is commonly
sampled from the carotid artery~\cite{Convert2022}, but the surgical procedure
for cannulation is terminal, making long-term studies on mice unfeasible. These
limitations highlight the need to explore non-invasive alternatives to arterial
cannulation.

Several advancements have been made in non-invasive estimation methods, such as
the population-based input function (PBIF,~\cite{takikawa1993PBIF}), or the
image-derived input function (IDIF,~\cite{henze1983IDIF} which uses TACs over
delineated arteries or blood pools, such as the left ventricle, in place of the
AIF~\cite{choi1991IDIFheart}, but demands rigorous calibration efforts to avoid
\textit{partial volume effects}, respiratory and cardiac motion, limiting its
practicality~\cite{Laforest2005,zanotti2011IDIFchallenges,frouin2002correction,kim2013partial,fang2008spillover}.
More recent efforts have investigated the use of machine learning-based input
functions (MLIF)~\cite{kuttner2020mlif,Kuttner2021cerebral15o}, but are limited
in needing manual TAC delineations.

\subsection{Deep learning derived input function}

This work builds on the foundation of recent advancements in deep learning-based
input function (DLIF) estimation methods for the
AIF~\cite{wang2024non,kuttner2024deep,ferrante2024physically}. Deep
learning-based methods have the advantage of processing the entire dPET data
volume, and can in principle learn mappings to mimic the IDIF, including
necessary calibrations. However, network design choices in how spatial and
temporal features are combined are still an actively researched topic.

In our previous work, proposing the FCDLIF~\cite{salomonsen2025DLIF} we combined
a spatial and temporal feature extractor to efficiently predict the AIF from the
dPET images of mice. The spatial feature extractor used the same 3D
convolutional network on each time frame to extract a dense representation which
the temporal feature extractor used to infer the values of the AIF. This
approach allowed an end-to-end pipeline that produced accurate estimates only
from the dPET images.

Wang \emph{et al.}~\cite{wang2024non} proposed a physics constrained AIF prediction
method, which given a set of regions-of-interest (ROIs) used a
LSTM~\cite{hochreiter1997LSTM1} feature extractor combined with
a fully connected prediction head to predict the input function. During
training, their LSTM-FC method used the loss from the residuals when fitting an
irreversible two-tissue compartment model (Figure~\ref{fig:2TC}, with $k_4 = 0$)
to each of their ROIs combined with the predicted input function. Their method
uses two loss terms, first computing the L1 loss between their methods
prediction and the IDIF, then averaging the residuals from the curve fitting
with the manually extracted TACs.

In this work, we propose a method that combines the efficient predictive
FCDLIF~\cite{salomonsen2025DLIF} method with the kinetic modeling loss presented
by Wang \emph{et al.}~\cite{wang2024non}. The proposed method, physics-informed
deep learning-based input function (PIDLIF), is trained on the same full-body
mice dataset as the FCDLIF~\cite{salomonsen2025DLIF}, using similar physical
guidance as the LSTM-FC~\cite{wang2024non}, but with only two ROIs. The PIDLIF's
ability to accurately operate on heavily distorted images as a result of
movement during imaging is investigated by simulating blurred voxels over time,
as depicted in Figure~\ref{fig:motionblur-example}. This type of motion can be
encountered during imaging due to breathing, cardiac or sudden movement induced
blurring~\cite{kotasidis2018robustness}.

\subsection{Kinetic modeling in dPET}

Compartment models and their kinetic parameters provide a mathematical approach to
describe the tracer kinetics in the body. It provides information about the activity of
the cells and tissue on a voxel or region-of-interest (ROI) level. A common model is the
two-tissue compartment model, which balances the tracer kinetics between tissue, denoted
$C_T(t)$ and measured arterial blood concentration, $C_A(t)$. In Figure~\ref{fig:2TC} we
see a reversible two-tissue compartment model, where $C_T(t) = C_1(t) + C_2(t)$, and
compartments $C_1$ and $C_2$ are indistinguishable in the image, but can be modelled
using a set of differential equations relating transitions between compartments with
rate constants $K_1$, $k_2$, $k_3$, and $k_4$: 

\begin{equation}\label{eq:2TCM ODE}
    \begin{aligned}
        \frac{dC_1(t)}{dt} &= K_1 \cdot C_A(t) - (k_2 + k_3) \cdot C_1(t) \\
        \frac{dC_2(t)}{dt} &= k_3 \cdot C_1(t) - k_4 \cdot C_2(t) \\
    \end{aligned}
\end{equation}

\begin{figure}[t]
  \centering
  \includegraphics[width=\textwidth]{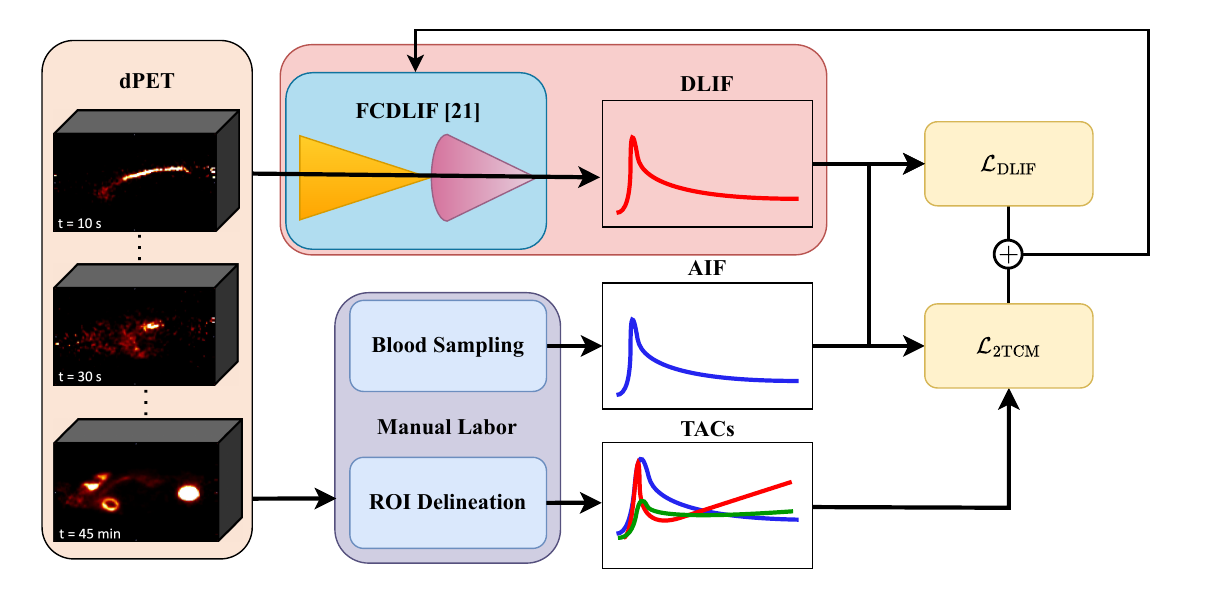}
  \caption{Our model in context: a time series of PET images are used as input
  to the spatial and temporal feature extractor from \cite{salomonsen2025DLIF},
  which predicts the input function. We compare this with the AIF using a mean
  squared error loss combined with the kinetic modeling loss, computed using a
  two-tissue compartment model, inspired by~\cite{wang2024non}.}
  \label{fig:model}
\end{figure}

Equation~\ref{eq:2TCM ODE} is generally simplified by assuming $k_4$ to be
negligible for shorter scan durations, ending up with the irreversible
two-tissue compartment model with solution derived using the \textit{Laplace
transformation} in the form:

\begin{equation}\label{eq:2TCM}
  C_T(t) = \frac{K_1}{k_2 + k_3} [k_3 + k_2 \cdot e^{-(k_2 + k_3)t}] \otimes C_A(t)
\end{equation}

The dashed line around the compartments in Figure~\ref{fig:2TC}, marked as
$C_\mathrm{PET}(t)$ describes what is visible in the dPET scan, where both a component
of the arterial blood and tissue is present in the image, and described using,

\begin{equation}\label{eq:PET Vb}
  C_\mathrm{PET}(t) = V_b \cdot C_A(t) + (1 - V_b) \cdot C_T(t),
\end{equation}

where $V_b$ is the blood volume fraction in the tissue. In compartment
modeling, the aim is to estimate the kinetic parameters by fitting the
parametrized curve $\hat C_T(t; K_1, k_2, k_3, k_4, V_b)$, to the curve $C_T(t)$
gained from a ROI segmentation that is averaged over time, or directly from the
spatial voxels. Once these parameters are determined, the net influx parameter
$K_i$ is computed as

\begin{equation}\label{eq:Ki}
  K_i \equiv \frac{K_1 \cdot k_3}{k_2 + k_3},
\end{equation}

providing a quantitative measure of the tracer's net uptake into the tissue.
Since $K_i$ is proportional to the metabolic rate of glucose, the resulting
parametric map significantly enhances lesion detectability. This improved
sensitivity facilitates the identification of small malignant tumours by
reducing the background signal from surrounding healthy tissue compared to the
original image.

\begin{figure}[t]
  \centering
  \includegraphics[width=\linewidth]{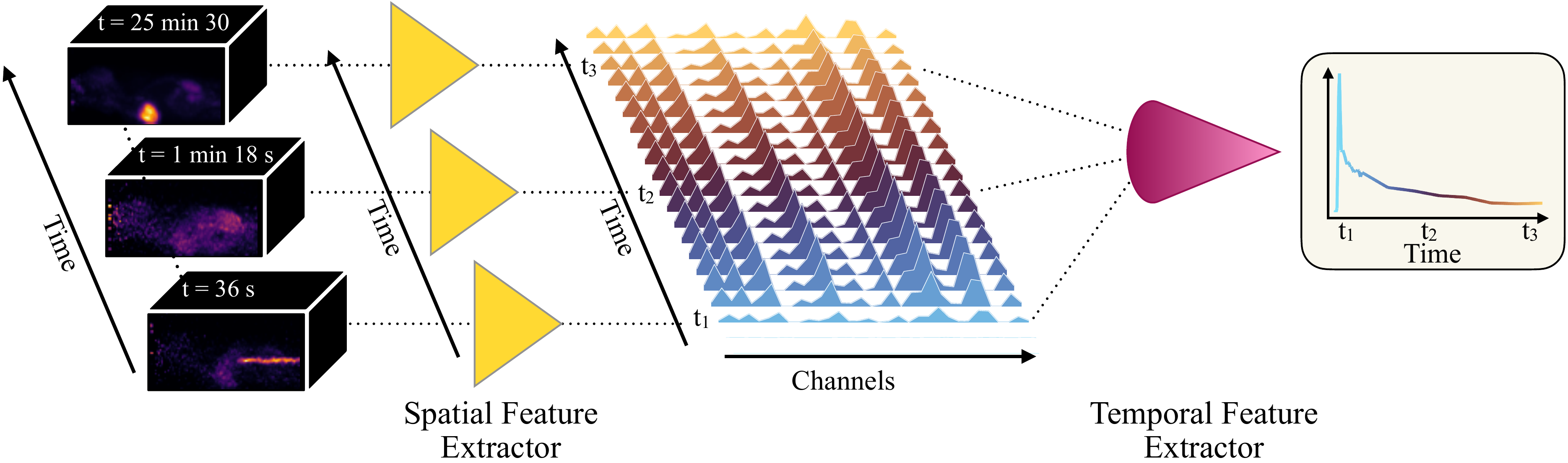}
  \caption{Input function prediction using FCDLIF~\cite{salomonsen2025DLIF}: a
  spatial feature extractor applies 3D convolutions in a bottleneck
  configuration to extract a dense representation of spatial features. This is
  followed by a temporal feature extractor that captures temporal relationships
  between adjacent time frames. The final output is the regression curve
  predicting the input function for the associated PET-images.}
  \label{fig:FCDLIF}
\end{figure}

\section{Methodology}

The proposed method sits on top of the recently proposed FCDLIF
model~\cite{salomonsen2025DLIF}, which is capable of directly predicting the
input function from a time-series of PET images. The model uses a two-step
approach, first extracting spatial features, sequentially applying 3D
convolutions to distill a dense representation. The second part is a temporal
feature extractor, and maps the latent representation to the different parts of
the predicted curve, as shown in Figure~\ref{fig:FCDLIF}. We refer the
interested reader to~\cite{salomonsen2025DLIF} for further details regarding the
FCDLIF model.

\subsection{Physics-informed loss function}

FCDLIF was trained using a weighted mean squared error (MSE) loss function,
where the tail of the predicted curves were more heavily penalized. The proposed
method compound this with a compartment modeling loss function, inspired
by~\cite{wang2024non}. To derive the physics-informed loss
$\mathcal{L}_\mathrm{2TCM}$, an irreversible two-tissue compartment model is
fitted to the brain and myocardium regions (delineated from $C_\mathrm{PET}$)
using both the predicted and measured input functions. Thus, each image is
fitted to two distinct compartment models, two times---one per input
function. The curve fitting process is described as,

\begin{equation}\label{eq:2TCM curve fit}
  \begin{aligned}
    \Phi &= \frac1n \sum_{i=1}^n \left(C_\mathrm{PET}(t_i) - \hat C_T(t_i;\hat p)\right)^2,\\
    \hat p^* &= \argmin_{\hat p} \Phi(\hat p),
  \end{aligned}
\end{equation}

where $n$ is the number of time frames in the spatio-temporal image, and $t_i$
is the time of the $i$-th frame. $\hat p^*$ is found by minimizing $\Phi$ using a
non-linear least squares method, and contains the rate constants $K_1,\ k_2,\
k_3$, and the fractional component of blood volume in tissue $V_b$. The proposed
loss function is then defined as the difference between the fitted compartment
model computed using the DLIF, compared with the corresponding compartment model
fit that used the AIF for the same ROI:

\begin{equation}\label{eq:2TCM loss}
  \mathcal{L}_\mathrm{2TCM} = \frac 1n \sum_{i=1}^n \left( \hat C_{T, \mathrm{DLIF}}(t_i;\hat p^*_\mathrm{DLIF}) - \hat C_{T, \mathrm{AIF}}(t_i; \hat p^*_\mathrm{AIF}) \right)^2
\end{equation}

In the above equation, $\hat C_{T, \mathrm{DLIF}}$ and $\hat C_{T,
\mathrm{AIF}}$ are the fitted curves using the DLIF and AIF, respectively. The
loss function is then combined with FCDLIF's loss function,

\begin{equation}\label{eq:total loss}
  \mathcal{L} = \mathcal{L}_\mathrm{DLIF} + \lambda \mathcal{L}_\mathrm{2TCM},
\end{equation}

where $\mathcal{L}_\mathrm{DLIF}$ is the weighted MSE loss function used in
FCDLIF. The total loss is then used to update the weights of the FCDLIF model
during training.

\subsection{Data}

The model is trained on a set of 70 dynamic [$^{18}$F]FDG PET/CT scans of mice,
accompanied by reference AIFs during imaging. The mice strain was BALB/cJRj
mice, at ages between 8 and 24 weeks. All mice were administered an intravenous
radiotracer dose of 120 ul over 30 seconds (240 ul/min), and the PET scan was
started 30 seconds prior to injection. The resulting spatio-temporal image per
scan and mouse is of the dimensions $42 \times 96 \times 48 \times 48$, where
each time frame was captured in a systematic ordering of $1\times 30s$,
$24\times 5s$, $9\times 20s$, and $8\times 300s$.

\subsection{Training procedure}

\begin{figure}[t]
  \centering
  \includegraphics[width=\linewidth]{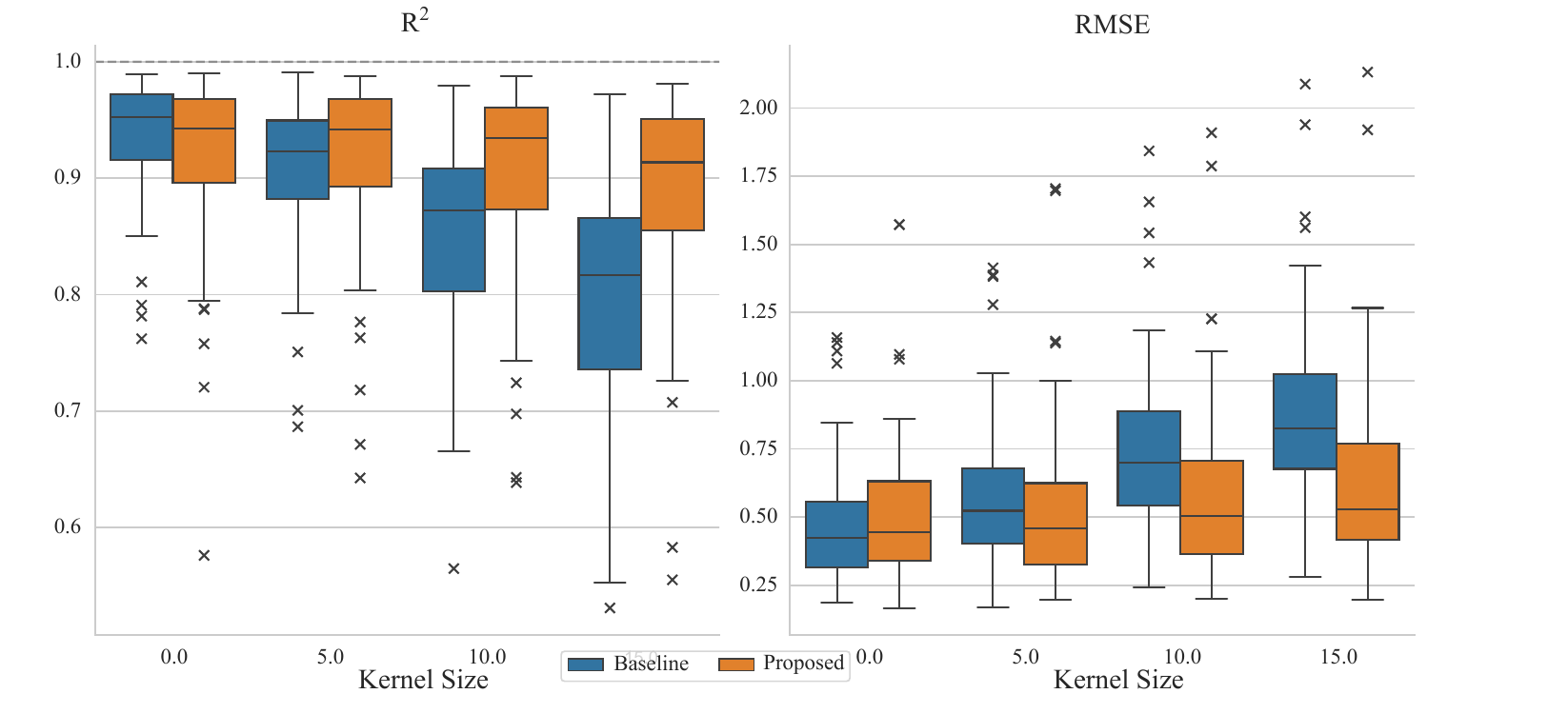}
  \caption{Results comparing the robustness of the
  FCDLIF~\cite{salomonsen2025DLIF} model with and without the physics-informed
  loss function to motion blur. The RMSE and $R^2$ values are shown for
  different levels of motion blurring denoted by the size of the kernel being
  convolved with the images.}
  \label{fig:results}
\end{figure}

For direct comparison with~\cite{kuttner2024deep}, we followed their training
procedure. Using the Adam optimizer~\cite{reddi2018on} with standard settings,
we trained the model with a learning rate of $10^{-4}$, and a batch size of 8.
The model is trained for 1000 epochs using 10-fold cross validation with 10 runs
within each fold, meaning each sample appears in the test set once. 15\% of the
remaining samples were set aside for validation during training. To account for
the imbalance in the sampling times of the curves, the peak, intermediate and
tail time steps of the input functions were associated with weights 0.4, 0.7,
and 1 respectively in $\mathcal{L}_\mathrm{DLIF}$. Since dPET images are
inherently noisy, Poisson noise was also introduced as a data augmentation
technique during training (see~\cite{salomonsen2025DLIF} for further details).
Due to instabilities with parameter optimizations of our physics-informed loss
during early epochs, when the input function predicitons are unreliable,
$\mathcal{L}_{2TCM}$ was gradually phased in by linearly increasing $\lambda$
from 0 to 1, between epochs 15 to 30.

\subsection{Evaluation of robustness}

To evaluate the models robustness to motion blur, movement is simulated to
degrade the quality of the image. This motion causes blurred voxels that further
amplify partial volume effects, where voxels smear into surrounding regions,
occluding the original signal~\cite{fang2008spillover}. This effect is shown in
Figure~\ref{fig:motionblur-example}, where for instance in the motion blurred
peak frame, the bright vertical \textit{Vena Cava} that runs from the bottom of
the image to the myocardium (bright region in the center of the image) bleeds
into nearby organs.

Practically, motion blurring is applied by convolving a 3D blurring kernel with
a randomly initialized direction. The motion is applied to each time frame, to
simulate persistent motion-induced blurring over time. To examine the variation
over different levels of blurring, the size of the blurring kernel is increased
in intervals of 5, up to 15 voxels over all dimensions.

\section{Results}

\begin{table}[t]
  \centering
  \caption{Evaluation results and comparison of increasing levels of motion blur
  induced image degradation. The kernel size refers to the size of the motion
  blurring kernel which is convolved with the images to simulate motion.
  Differences between the baseline~\cite{salomonsen2025DLIF} and the proposed
  method were assessed by two‐sided paired $t$‐tests; entries with $p<0.001$
  indicate a statistically significant difference (i.e. the baseline and
  proposed scores differ significantly), and are marked in bold. The arrows
  indicate whether an increase or decrease in the metric is better.}
  \begin{tabular}{lc|c@{\hskip 0.3cm}c}
    \hline
    Model & Kernel size & $R^2$ $\uparrow$ & RMSE $\downarrow$ \\
    \hline \hline
    Baseline & 0 & $0.94 \pm 0.05$ & $0.47 \pm 0.22$ \\
    Proposed & 0 & $0.92 \pm 0.07$ & $0.51 \pm 0.27$ \\
    \hline
    Baseline & 5 & $0.91 \pm 0.06$ & $0.58 \pm 0.27$ \\
    Proposed & 5 & $0.92 \pm 0.07$ & $0.53 \pm 0.30$ \\
    \hline
    Baseline & 10 & $\mathbf{0.85 \pm 0.09}$ & $\mathbf{0.74 \pm 0.31}$ \\
    Proposed & 10 & $\mathbf{0.90 \pm 0.08}$ & $\mathbf{0.58 \pm 0.33}$ \\
    \hline
    Baseline & 15 & $\mathbf{0.79 \pm 0.11}$ & $\mathbf{0.88 \pm 0.34}$ \\
    Proposed & 15 & $\mathbf{0.89 \pm 0.09}$ & $\mathbf{0.64 \pm 0.35}$ \\
    \hline \hline
  \end{tabular}
  \label{tab:results}
\end{table}

We evaluate our model using the root mean squared error (RMSE) and the
coefficient of determination ($R^2$) between the predicted and true AIFs.
Figure~\ref{fig:results} and Table~\ref{tab:results} with a kernel size of 0
reports the base performance on the original images. From these results, the
slight decrease in $R^2$ and RMSE for the proposed method demonstrates that the
physics-informed model has similar performance to the
FCDLIF~\cite{salomonsen2025DLIF}, but does not outright surpass it on the
noiseless data. This is expected as the FCDLIF is already capable of accurate
AIF estimation. The slight decrease in performance may be an artefact of the
longer convergence time when the additional physics-informed loss is involved,
occasionally failing to converge in the given training time. This indicates the
learning objective during training increased in complexity with the addition of
the physics guidance.

In Figure~\ref{fig:results}, where the kernel size is greater than 0, the
physics-informed DLIF's robustness to motion blur can be observed. The RMSE and
$R^2$ values are shown for different levels of motion blur, where the PIDLIF is
not as affected by the increased degradation as the
FCDLIF~\cite{salomonsen2025DLIF}. From Table~\ref{tab:results} (kernel size $>
0$), the physics-informed model's RMSE increases by 25.4\%, compared to an
87.2\% increase in the FCDLIF~\cite{salomonsen2025DLIF}, from no blur to the
maximum amount of blur added. Likewise, the $R^2$ scores report a similar trend
where at the highest levels of degradation, the 75$^{th}$ percentile of the
proposed model is nearly at the 25$^{th}$ percentile of the adjacent FCDLIF
distribution.

\section{Discussion}

This study introduces PIDLIF as a physically guided method that uses an
irreversible two-tissue compartment model during the training phase to ground
the predictions of the network. The approach uses the predictive framework
from~\cite{salomonsen2025DLIF}, borrowing ideas from Wang \emph{et al.} on the
physics-guidance. Through online compartment modeling of ROIs during training,
both the FCDLIF~\cite{salomonsen2025DLIF}, and the LSTM-FC
from~\cite{wang2024non} adds additional regularization in their prediction
objectives.

Our PIDLIF framework differs from the LSTM–FC DLIF model of Wang \emph{et al.}
in several key aspects. First, whereas Wang \emph{et al.} ingest manually
extracted TACs from eight cerebral regions plus the carotid artery and employ a
three‐layer LSTM followed by a two‐layer fully connected prediction head, PIDLIF
is an end‐to‐end convolutional network that maps the full 4D PET volume directly
to an AIF estimate~\cite{salomonsen2025DLIF}. This reduces reliance on
labor‐intensive ROI delineation, enabling direct prediction from imaging data.

Second, the two physics‐informed losses differ. Wang \emph{et al.} integrate an
adaptive weight $\lambda_k$, computed every five epochs based on the normalized
change rates of the input function and kinetic modeling losses, so that the
physical constraint gradually equilibrates with the MSE loss over
training~\cite{wang2024non}. In contrast, in PIDLIF the physics-informed term is
linearly faded in between epochs 15 (zero attribution in loss) and 30 (equal
attribution in loss), simplifying hyperparameter tuning and stabilizing early
convergence. Moreover, Wang et al. penalize the residuals from curve‐fits of
each ROI TAC against the predicted input function, that is, $\hat C_T(t;\hat
C_A,\hat p) - C_\mathrm{PET}$, where $\hat C_A$ is the predicted input
function. However, PIDLIF minimizes the difference between the two-tissue
compartment model fits obtained with the predicted and ground-truth AIF
(Equation~\ref{eq:2TCM loss}, to simplify the objective function, as there is
usually some discrepancies between the image-based ROI measurements and the
fitted curve, $\hat C_T$.

\subsection{Physical guidance improves robustness to motion blur}

Across modest data augmentations, e.g. Poisson noise, small movements already in
the data, PIDLIF and the baseline FCDLIF achieve comparable AIF accuracy,
indicating that purely data‐driven convolutional features suffice when image
degradation is limited. However, under simulated motion‐blur artifacts,
mimicking movement, PIDLIF retains substantially higher fidelity (e.g.
<30\% RMSE increase vs. >80\% for FCDLIF at maximum blur), and maintains
superior $R^2$ scores (Table~\ref{tab:results}). This suggests that
incorporating kinetic modeling constraints enhances out‐of‐distribution
robustness by enforcing physiological
consistency~\cite{Karniadakis2021,banerjee2024pinnsmedicalimageanalysis}.
Nonetheless, in the absence of severe distortions, the baseline convolutional
model slightly outperforms PIDLIF in raw regression error, reflecting the extra
optimization complexity introduced by the physics term.

\subsection{Limitations}

While PIDLIF shows promise, several limitations warrant discussion. Firstly, the
mouse cohort (70 scans) is large by small-animal study standards, but small for
deep learning, requiring extensive cross-validation efforts for statistical
rigour~\cite{salomonsen2025DLIF}. The physics-informed loss is constrained to
two ROIs (myocardium and brain) for simplicity, and since the organs adhere to
the two-tissue compartment assumptions. Extending the physics-informed loss to
several ROIs, or full image voxel-wise compartment models could better help the
networks learn more robust representations, but convential non-linear
optimization strategies are too computationally expensive for a full PET
alternative.

The simulated blurring kernel only models persistent voxel smearing; real motion
can be abrupt, non-rigid, or intermittent, and can also be thought to degrade
the attenuation correction. These effects are not captured in our simplified
experiments, and requires more advanced simulation techniques.

Another limitation relates to the added complexity during training. Since the
physics-informed term substantially increases optimization difficulty,
occasionally slowing convergence, more adaptive or multi-stage training
schedules may be needed as a surrogate for more data samples.

\section{Conclusion}

In conclusion, PIDLIF demonstrates that embedding tracer‐kinetic constraints
into an end‐to‐end convolutional AIF predictor can markedly improve robustness
to severe image degradation from blurring, at the expense of increases in
training complexity. Future work will extend the physics-informed loss to
full-PET voxel-wise compartment modeling, and the use of unsupervised
clustering algorithms for automatic ROI delineations, which might offer a
steadier alternative to the high-variability associated with voxel-wise
computations.

\bibliographystyle{splncs04}
\bibliography{bibliography}

\end{document}